\documentclass[letterpaper]{article}

\usepackage[english]{babel}
\usepackage[utf8x]{inputenc}
\usepackage[T1]{fontenc}
\usepackage[stable]{footmisc}

\usepackage[letterpaper,margin=1.0in]{geometry}

\usepackage{amsmath}
\usepackage{graphicx}
\usepackage[colorlinks=true, allcolors=blue]{hyperref}

\title{Networking Research – A Reflection in the Middle Years}
\author{Henning Schulzrinne\\Columbia University}

\begin{document}
\maketitle

\begin{abstract}
Networking is no longer a new area of computer science and engineering -- it has matured as a discipline and the major infrastructure it supports, the Internet, is long past being primarily a research artifact. I believe that we should consider ourselves as the civil engineers of the Internet, primarily helping to understand and improve a vast and critical infrastructure. This implies that implementing changes takes decades, not conference cycles, and that implementation is largely driven by compatibility with existing infrastructure and considerations of cost effectiveness, where resources that research focuses on, such as bandwidth and compute cycles, often play a much smaller role than limited organizational capacity for change. Telecommunications carriers, in particular, have become akin to airlines, largely operating equipment designed by others, with emphasis on marketing, not innovation. Even more than in other engineering disciplines, standards matter, whether set by standards bodies or dominant players. Given the multi-year time frames of standards and the limited willingness of national funding bodies to support standardization work, this makes research impact harder, as does the increasing complexity of cellular networks and barriers to entry that shut out most researchers from contributing to large parts of commercial mobile networks.
\end{abstract}

\section{Introduction}

We’re no longer young – 30th (web, in 2019), 40th (Computer Networks) and 37th (IEEE Infocom and TCP) anniversaries of technology milestones, journals and conferences make it plain that packet-based and Internet technologies are now well into their family and mid-career stages, where promise and potential are replaced by concerns about paying technology debts, legacy (systems) and thinking about retirement, for both legacy systems and some of the pioneers of the field.
Much of what I will say in this piece is not true as stated – there are exceptions, many of them, and I hope others will point them out, but ``on the one hand, on the other'' writing is not all that interesting and misses the fact that these hands are not ambidextrous. Also, my curmudgeon take is not meant to denigrate the good work, often with great patience and diligence, the networking community has done over the past decades. But there are plenty of keynotes and award acceptance speeches celebrating those.

\section{We are Civil Engineers Repaving Roads}

Communication networks, whether telephony and telegraph through the 1980s, or data networks since then are foundational components of any economy, along with water, electricity (and other energy supply chains) and transportation. Indeed, they are probably foundational in the sense that energy and other services can be improvised locally to some extent after natural disasters, but without communication networks, it is hard to coordinate recovery, restore commerce or reduce the anxiety of loved ones.

Like other widely-distributed, interconnected infrastructures that represent hundreds of billions of dollars of investment even within one country\footnote{The total asset value of the 30 largest telecommunications providers is estimated at \$2.4 trillion \cite{Liu14:Telecom}.}, communication networks change slowly. Indeed, it can be said that we have had only three generations of (electric) communication networks – telegraph, telephone and the Internet, each taking roughly half a century from inception to near-universal availability. The first two were limited to either text or (mostly) voice, but since the Internet can transport any information content expressible in bits, it seems likely that this last iteration will last until quantum transport becomes viable, just like motorized vehicles still look and function in ways that the owner of a 1908 Ford Model T would recognize and a Boeing 707 airliner from 1958 looks familiar to a 2018 pilot. (We will get autonomous versions of networks, too.)

Just like many civil engineering text books from the 1990s are probably still in slightly revised editions, networking textbooks from the same period are still a good description of how commercial networks operate \cite{Kuro99:Computer,Stev9312:TCP,Tane81:Computer}, with ATM, FDDI, DQDB, and WiMax making brief transient appearances on the introductory networking class stage.

\section{It Always Takes Longer than You Think}

There is a conceit that networking is a fast-moving area, with ``rapid advances'' being the cliché introducing many a conference paper. But deployment of new technologies takes extraordinarily long, measured in decades, for reasons that are somewhat unique to networking: First, scale matters more than efficiency or functionality. Thus, being able to connect to legacy technologies has customer and financial value. New applications are often hard to predict at all, and harder to predict in their impact. After all, 3G and 4G were premised on supporting IMS for multimedia calls, but succeeded because of web browsing. Many of these predictions, from peer-to-peer and 3D TV to IoT and VR, also turned out to be wildly optimistic in timing, relevance or longevity.

This is one reason why many traditional telecom vendors, and their carrier customers, struggled to adapt to the transition to IP and VoIP, as commercial data services and SS7-based voice supported their existing customers quite well and quite profitably\footnote{This would be the usual place for the obligatory reference to the \emph{Innovator's Dilemma} \cite{Chri97:Innovators}, but there is sufficient doubt about its empirical foundation that I will refrain.}.

Examples are well known: VoIP standardization was largely completed in the mid-1990s, but TDM voice still commands roughly half the (shrinking) wireline market and VoLTE is just getting deployed universally, 20 years later. Much of carrier-to-carrier interconnection is still TDM with SS7 signaling. IPv6 was published in 1995 \cite{rfc1883}, but is just now reaching about 19\% of Google traffic\footnote{Google IPv6 statistics at https://www.google.com/intl/en/ipv6/statistics.html}. Many public services, such as public safety communications, are still using FM trunk radio and digital circuits designed for long-distance telephone operators\footnote{CAMA trunks for 9-1-1}. We still cannot prevent unwanted robocalls.

Secondly, with a few exceptions, the capabilities and efficiency of today’s networking technology are not all that different from those in the 1990s. Clearly, speeds have increased by orders of magnitude, but largely due to much-improved physical layers and faster electronics, not advances in congestion control, routing or application-layer protocols. We are encrypting many of our network protocols, but it is not obvious that confidentiality, integrity and availability have improved significantly.

\section{Publish It and They will Implement}

We need to think of ourselves as an engineering discipline — we are not providing insights into fundamental laws of nature\footnote{And when we try, as for notions of the universality of power law networks, we tend to over-generalize} or the fundamental nature of humans. Thus, our societal value is found in improving a core civilizational infrastructure and we should be measured on that score, not on citations or mathematical sophistication.

Unfortunately, we have had somewhat gauzy notions of how research and publications transition into practice and rarely seem to actually study that transition. I suspect all of us believe at some level that our brilliant idea that just got published in a highly-selective conference will be studied by engineers at a major manufacturer with awe and somehow become part of an industry-defining product. In some cases, the transition can be traced fairly easily, just by looking at who contributed to both research and standards, but in many other cases, even if ideas that were found in academic publications later end up in standards, products and deployment, it is not clear that the publication itself caused this to happen or whether this was just the natural solution that a competent practicing engineer who never consulted IEEE Xplore or the ACM Digital Library came up with.

Conversely, most ideas published even in the most selective conferences never have any impact at all, beyond citations in other papers that unfortunately share the same fate. Despite all of the collective effort of program committees and journal editors, this is probably unavoidable. Given that a good idea implemented can add billions of dollars of value, having thousands that turn out to be interesting, but never see deployment, is not a bad bet, particularly given the delay between idea and implementation that would never let any student graduate or any faculty get tenure. Also, it often takes many small steps to get to a sufficiently ``baked'' idea — standing on the shoulders of lots of midgets, as it were. Unfortunately, the re-usability of the median networking research is somewhat lower than in some other fields; even modest medical studies can become crucial parts of larger meta-studies, and the basic biology of humans does not change all that much, so they remain relevant for decades. Now that we have fifty years of history, it would be extremely interesting to go through core components of the deployed Internet and see if their intellectual lineage can be traced to papers or research efforts.

However, as we overestimate the impact of conferences and journals, we underestimate the inter-generational transmission of technology advances and culture. At many a faculty retirement party, the emeritus-to-be notes that his or her students were his favorite output. They probably also had far more impact than their MS or PhD thesis, acting as a means to spread ideas and approaches by social ``contagion''. For those graduates, any published paper probably had most of its impact as a training exercise in good engineering, careful evaluation and convincing presentation. A classic example is the distinction of ``Bell heads'' from “net heads”, which, at least partially can be explained by their academic and early-career upbringing. Thus, thinking carefully what principles and ideas we teach the next generation may be more important than the papers we publish.

\section{The Streetlight Effect\footnote{See https://en.wikipedia.org/wiki/Streetlight\_effect}}

Until recently, almost all academic layer-2-and-above work published seemed to focus on Wi-Fi, with very little in-depth work on LTE and other cellular networks. The reason seems fairly simple: any single-faculty lab could buy Wi-Fi equipment, both base stations and clients, and run experiments without worrying about having the FCC or equivalent local regulatory agency shut down the lab for causing RF interference with licensed commercial services. Even testbeds like ORBIT\footnote{See http://www.orbit-lab.org/} initially mostly supported Wi-Fi nodes, with attempts at WiMax experiments having less success\footnote{We ran a WiMax node on the Columbia University campus, after spending a year trying to get facilities to install an antenna on a university building and significant effort to get an experimental license.}. Besides the still-high cost of equipment and access to spectrum for experiments, the complexity of current cellular systems makes them ill-suited for academic research. A student will be ready to graduate, and be a CS and EE double major, before they understand all the details of LTE RANs.

\section{Agencies Want Broader Impact, but Panels Won’t Fund Standards Work}

Many a discussion at NSF panel review sections has discussed broader impact\footnote{https://broaderimpacts.net/}. Often, this is simply a short-hand for running a graduate seminar addressing the topic of the proposal, but translation to industrial practice through standards is not a foreign concept. However, proposals that focus on having faculty participate in standardization or address topics directly related to standards are often seen as lacking that other vital ingredient, intellectual merit. Since everybody knows that, such proposals are also rare. Not surprisingly, the number of faculty (and graduate students) who regularly participate in Internet or IEEE 802 standardization is tiny.

There are other reasons for this lack of interaction, with the very different timelines one concern. Often, even small-bore standards can take two years, or more, to complete, with maybe one paper to show for it. This is a high-risk approach for faculty watching the tenure clock.

\section{Math Hammers, Looking for Nails}

One of reasons that QoS has been such a consistent theme in networking is that it offers a natural opportunity for applying new mathematical tools, whether initially queuing theory or, most recently, machine learning. Often, these problems then lead to advances in those tools, so there is something to be said for the approach, even if they do not directly lead to deployed systems.

Since every new networking technology poses familiar, yet sufficiently new problems, every new networking technology attracts a “QoS for \ldots” research effort, whether that is Wi-Fi, sensor networks, ad-hoc networks, peer-to-peer networks, cloud computing or ICN. I am still waiting for the QoS for blockchain paper, but I’m sure it’s being worked on. (Blockchain for QoS presumably has been published already \cite{Sing17:Comparative}.)

\section{Telecommunication Carriers are Like Airlines}

We tend to think of telecommunication carriers, whether broadband Internet access or backbone providers, as technology companies. This was not unjustified during the era when AT\&T Bell Labs and its smaller cousins operated by carriers worldwide dominated industrial research across wide swaths of communication and computing, but that era ended roughly twenty years ago. Now, I think it is more useful to consider carriers as the rough equivalent of airlines. Just like airlines, they buy highly complex equipment from a small number of equipment vendors, often influencing the design and requirements, with an emphasis on longevity and reliability. But they all operate the same equipment, often for decades with incremental updates, distinguished largely by the color scheme of their advertising, their pricing model and their own-time performance\footnote{Unfortunately, airlines and ISPs, at least in the United States, also share the distinction of ranking among the least-loved businesses. According to the American Customer Satisfaction Index (ACSI)\cite{ACSI}, the telecommunication industry ranks lowest, ``beating'' the local Department of Motor Vehicles and U.S. airlines by a significant margin.}.

Carriers have forever wanted to be more than dumb bit pipes, but have never really succeeded in offering competitive services above the network layer. (Some see the resistance of carriers to strong network neutrality rules as their attempt to give their vertically-integrated services a pricing or technology advantage, or to monetize their unique ability to reach their customers.)

As a commodity carrier, technology innovation rightfully becomes purely a financial cost-benefit analysis. There just is not a whole lot of incentive to rip out existing TDM voice when this is seen by carriers as a cash flow business similar to offering dial-up Internet service or when it is a zero-revenue add-on service. Similarly, deploying IPv6 is not going to happen unless the economic pain of acquiring IPv4 addresses or kludges like carrier-grade NATs become overwhelming.

\section{Bits, Bytes and Cycles are Cheaper than Humans}

For almost the entire history of networking research, the only resource that was explicitly being optimized for were first bits on the wire, then CPU cycles and memory and, more recently, energy. However, most carriers only spend about 15\% of their revenue on capital investment\cite{Liu14:Telecom}. For new network builds or upgrades, almost all of that goes towards civil engineering and fiber, not routers or software. Unfortunately, many of the means to improve efficiency also increase operational cost, by increased complexity, and the  need for more highly-trained operations staff\footnote{It has been said that PIM-SM was designed by PhDs and required a PhD to understand and operate.}, and may make the network more brittle, e.g., as headroom is reduced, cross-layer or cross-element dependencies are introduced or mis-configurations are made more likely\cite{FCC1705:March}.

Interestingly, judging from industry publications like LightReading\footnote{LightReading; https://www.lightreading.com/automation.asp}, almost all the interest at carriers, outside of 5G, has moved to improving operational efficiency, by automating the boring parts that attract almost no attention even in network management conferences, such as ordering new service or self-diagnosing failures towards the edge. In other words, increasing traffic loading by some fraction is far less productive than shutting down a customer call center, as unfortunate that is for the staff working there. SDN and NFV are mostly not about new functionality, but attempts to reduce the need for humans to touch network settings.

Unfortunately, and possibly unavoidably due to lack of data or willingness of carriers to reveal data, these types of economic cost-benefit analyses, common in other areas of engineering, have been missing from networking research, leaving a gaping hole for relevance and comparison.

\section{Statements of Faith}

For reasons that I do not claim to fully understand, networking seems to attract religious fervor, comparably only to older-days preferences between emacs and vi, but with more at stake. Examples, both past and current, include OSI networks in the early days, ATM\footnote{For younger readers, this was the cell-based transport mechanism meant to initially create the broadband ISDN, as a better replacement –- QoS! –- for the Internet.} in the 1990s, sensor networks, ICN and, most recently, 5G or, further afield, blockchains. Other buzzwords, like LTE IMS or DQDB, attracted smaller communities of dedicated followers.

There are probably a few mechanisms at work, without claiming to have any data or experiments to support these hypotheses. Packets are not all that exciting by themselves, so engineers may look for causes to believe in or tribes to belong to. If one is looking for funding, it is not harmful if everybody else, in particular leaders of funding agencies, can be made to believe that this technology is the key to the future, addressing the trifecta of QoS, security and societally-valuable applications.

Unfortunately, both engineers and non-engineers believe the hype, even though beer-enhanced dinner conversations may be more honest\footnote{Dagstuhl deserves a special call-out here.}, or cynical. I suspect that we have very few careful analyses why technologies that were seen as the next big thing underperformed expectations or slowly faded away. After all, ATM was the hot item at conferences and commercially successful, until it was not, but I am not aware of any in-depth analysis of what technology and economic factors led to the ``victory'' of other layer-2 or layer-3 technologies and whether the technology continues to live on indirectly in ideas or terminology. An outside observer comparing network textbook editions might be reminded of Stalinist purges where, without explanation, leading functionaries suddenly were no longer in the picture. Postmortems are useful for any large project, but we do not seem to have much interest in that, moving on to the next big thing.

\section{Industrial Research is a Monopoly Game}

Networking research, probably more than other parts of electrical engineering and computer science except maybe computer architecture and VLSI, has always relied heavily on the contributions of both university and industry labs. The latter could often work on larger projects with lots of ``boring'' parts, had closer connections to the standardization community and could more easily build systems that required new hardware and software. In the early 1990s, we had a very diverse eco system of industrial labs, at scales from dozens to thousands of researchers, with well-known labs like AT\&T Bell Labs, Bellcore, Cisco, DEC, Ericsson, Fujitsu, HP, IBM, Intel, Marconi, Microsoft, Nokia, Nortel, Siemens, Sun Microsystems, and Xerox PARC all producing first-rate and high-impact networking work, and with their researchers participating actively in technical program committees and the academically-dominated side of professional societies. Every major telecommunication carrier had one or more research labs that could be seen as a great destination for systems-oriented PhD students. As the equipment vendors have consolidated into three companies (Ericsson, Huawei and Nokia) and as carriers have largely abandoned work other than systems integration, down-stream standardization\footnote{For networking, I am thinking of standards bodies like ATIS and 3GPP that largely, in layers 2 and above, standardize profiles of standards developed elsewhere, not algorithms and protocols.} and testing, the number of labs with external visibility has gotten much smaller.

\begin{figure}
\centering
\includegraphics[width=0.8\textwidth]{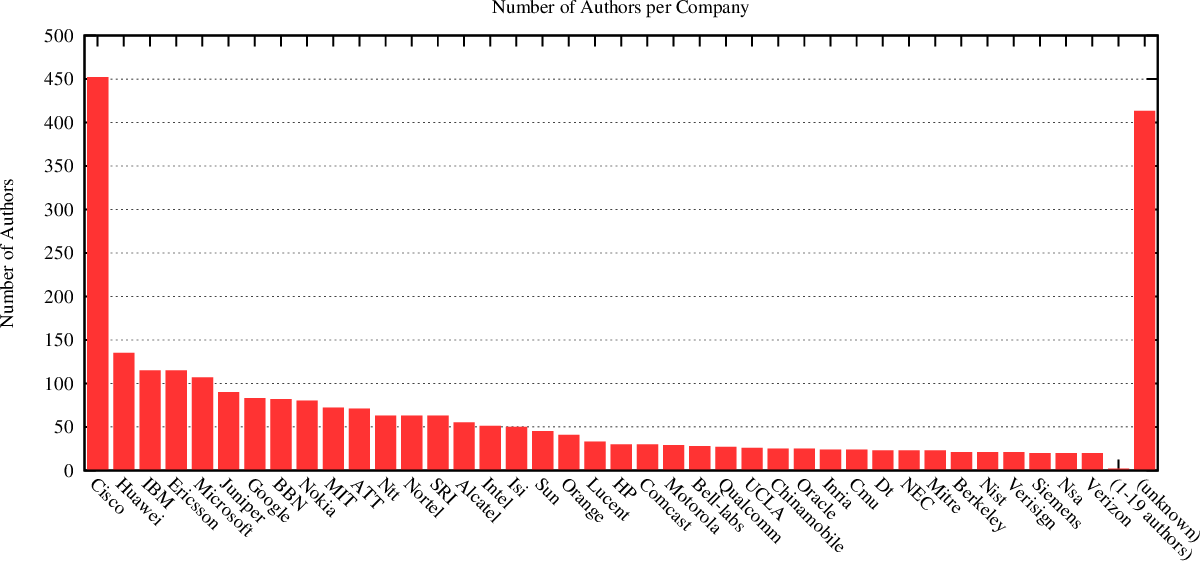}
\caption{\label{fig:ietf}Number of RFC authors from different organizations \cite{Arkk:Company}}
\end{figure}

It has been observed that industrial research works best with a hungry monopolist, i.e., an entity that can capture a significant fraction of the economic benefit (positive externalities) of research without having to rely on patents or trade secrets, and also wants to expand into new areas. Earlier monopolies or dominant companies also had immediate paths to commercialization at scale, driving adoption. Google, Microsoft and, in an earlier era, AT\&T Bell Labs and IBM fit this predictor. As networking technology as become more commoditized, this seems harder to find. Not surprisingly, Google and Facebook have become drivers of applied networking research, e.g., with Google’s work on QUIC, HTTP/2 and similar more significant departures from current, largely ossified, practice.
The list of contributors to IETF standards in Fig.~\ref{fig:ietf} illustrates the problem that most authors of RFCs are employed by a few companies. (Note that the figure does not reflect the number of RFCs, just the number of authors, so it does not reflect relative contributions.)

\section{Stuck in the Middle}

Networking research is largely defined as encompassing the link up through ``middleware'' parts of the application layer. The networking community has not had much impact on developing new applications, as applications such as the web, e-commerce, messaging, social media, streaming (HTTP) video, peer-to-peer, and blockchains seem to have largely arisen at the periphery. (In some of these cases, the key developers were aware of some of the related academic or industry work.) Similarly, almost all increases in speed and decreases in latency have been due to advances in optical, copper (DSL and HFC) and wireless physical layers (IEEE 802.11, LTE and its predecessors). Thus, networking research is, to some extent, bounded by what happens both below and above. This is hard to avoid, but probably explains some of the challenges the field will have to address in the future to maintain impact and relevance. Fortunately, as noted, even small advances in transport efficiency, routing stability or security can have a major impact. But, judging from recent conferences, much of the networking research has moved to the edge, except for work related to wireless links.

Convincing the next generation of students and the current generation of university administrators is going to be a challenge. For example, in 2015, 126 out 1297 PhD or 9.7\% chose operating system or networks as their dissertation topic,  vs. 13.7\% in 2001\footnote{Statistics by the Computing Research Association (CRA); OS and networking are treated as one area in the earlier years, so only aggregate comparisons are possible.}.

It would not be the first time that an infrastructure-related discipline has seen significant decreases in university investment. For example, power engineering gradually disappeared from electrical engineering departments in the United States, and there are few if any railroad, pipeline or shipping research centers at  universities there.

\section{Pleading Poverty}

Resource poverty is one of the most persistent themes in research, with a stock paper outline following the pattern: “Bandwidth (or CPU or memory or spectrum or IP addresses) is scarce, so let me propose a sophisticated algorithm to fix the problem for a narrow subset of applications, run some simulations in a toy environment or some analysis assuming Poisson arrivals and claim 10\% improvement for my carefully selected set of cases.” This, along with the availability of mathematical tools, explains the persistent interest across 40 or more years in all things related to quality of service, despite equally persistent lack of commercial success of all but the most basic, priority-based QoS mechanisms.

From a research perspective, the unfortunate fact has been that improvements in the cost of computation, storage and bandwidth have often outpaced the ability to deploy networking solutions. After all, deploying a new MAC layer in 802.11 that offers a 10\% improvement is harder than using MIMO to double or triple the throughput. IoT devices now have computational power exceeding that of a Sun workstation\footnote{\emph{Raspberry Pi vs SPARCstation 20: Fight!} at http://eschatologist.net/blog/?p=266}. Quality of service does not create capacity -- it just pushes aside the applications the writer of the paper considers less worthy.

A good example of this resource fixation was the near-decade long work on sensor networks. It is not clear that the work has had any major influence on its in-spirit successor, the Internet of Things, as they mostly seem to run standard operating systems, Wi-Fi and web protocols, plus MQTT, largely because systems-on-a-chip running Wi-Fi have become dirt cheap and most IoT applications have access to a power source since they control something attached to the electric grid or a vehicle.

\section{You Cannot Replicate Yourself to Funding or Tenure}

Other disciplines that have a strong experimental focus have started to worry about a replication crisis \cite{Stro14:Alleged}. There have been concerted and high-visibility efforts in, say, psychology to reproduce experimental results. While ACM and IEEE have started to look at reproducibility, e.g., through making data or code available\footnote{A countervailing trend is that one of the most promising ways to get published in a highly selective conference is to have exclusive access to a desirable industrial data set that is almost never shared.}, the appetite for actual reproducing other people’s work seems limited. As far as I can tell, conferences and journals do not publish such work, so the only real hope for encouraging such work is the unlikely chance that the original author made a major mistake. As long as p-hacking is not a promising approach in networking, this seems to limit the incentives. If reproducing earlier work is unlikely to happen just by providing data and code and if high-visibility work relies on ``secret'' data, we may want to reconsider the incentives.

\section{Resource Scarcity, the Money Part}

Networking research has been the beneficiary of the Internet halo for many years, probably attracting a disproportionate amount of funding in Europe, the US and Asia. In the US, both infrastructure and curiosity-driven research were funded by NSF, while DARPA funded larger projects. Industry labs contributed significant resources. For FY 2017, total core NSF network funding is only \$14.4M, with DARPA and industry funding having largely dried up. Thus, many future networking projects will probably have to find a home in application areas, where fortunately just about every technology area will rely on communication networks, as the “CS+X” equivalent.

\section{The Internet is Not Meant to be Secure}
\label{sec:security}

One of the standard conference panel discussion tropes is to state that the birth defect of the Internet is its lack of built-in security, such as the lack of encryption and authentication. But adding security features to lower-layer Internet protocols has only been modestly successful. It is not obvious why replacing TLS with IPsec would, for example, make the Internet substantially more secure. So far, as best as I can tell, TLS has not prevented any major credit card thefts nor hindered the activities of Cambridge Analytica nor made DDOS or ransomware less common. (Indeed, I would bet that if we turned off TLS, the number of pilfered credit card numbers would increase only marginally.)

Thus, we need to move beyond the well-worn bromide of making the Internet secure and explain how exactly this would work as long as end systems and routers run buggy software and as long as increasing complexity makes configuration mistakes likely. As long as the cost to perpetrator and enablers are low, a better Internet layer seems to offer only marginal improvement. (It does not help that known mechanisms such as source address validation (BCP 38, \cite{rfc2827}) do not get deployed universally, even after twenty years.)

\section{Connectivity Is Not Good}

Well into the 2010s, we could and did pretend that more, better, faster connectivity was always a good thing, without any qualifications. We looked down on legacy media like newspapers, telephony, radio and TV as dinosaurs that deserved to die, obliterated by the meteor of the Internet. As downsides like spam and cyber-attacks could no longer be denied, we could still attribute them to the abuse of our good intentions by a few evil people, and all we would need is a future Internet research program that fixed the foundational flaw of lack of built-in security (see Section~\ref{sec:security}). 

Rural broadband would fix the left-behind areas in Europe and the United States; telemedicine would dramatically improve health outcomes and efficiency; distance education and massive online courses would equalize educational attainment, while not having to pay those expensive teachers and college professors; inter-vehicle communication would dramatically reduce accidents and congestion; blogs and social media would turn everyone into an influencer; social media would enable civic society to overthrow tyranny; writing apps would provide secure jobs to millions. Peter Barlow declared the independence of cyberspace from nation states \cite{Barl9602:Declaration}.  All of these statements have a kernel of truth — clearly, if you want millenials to stay in rural areas, offering decent broadband is as necessary as having running water and electricity \cite{Gall1804:When}. Telemedicine can provide mental health counseling to veterans who might otherwise not be able to travel to the nearest VA hospital. Distance education and MOOCs can supplement educational offerings to dedicated learners. But in almost all cases, everybody had an interest to oversell the benefits and ignore the downsides or less-savory uses. For magazines like Wired, it offered the opportunity sell a new lifestyle and glossy ads. For researchers, it provides the necessary broader impact and public benefit; for politicians and their donors, it can be a cheap alternative to funding local schools, to paying teachers a living wage, to improve public transportation or to find sustainable economic models for rural communities. For Internet start-ups, it made regulation seem unnecessary and provided a halo obscuring the economic impacts on, say, local newspapers and main street stores.

There were other voices, with Cliff Stoll \cite{Stol96:Silicon} raising issues of functionality and usability that have been mostly addressed \cite{Bren1703:Web}, but mostly missing the larger societal impact. Tim Wu’s \emph{The Master Switch} \cite{Wu10:Master} deserves more attention in the engineering community as it at least identifies the common pattern of ascribing liberating, decentralizing and democratizing advances to new media, whether radio or TV or the Internet, which then invariably turns into technologies dominated by a relatively small number of large corporations, and from education to more profitable advertising-supported entertainment.

Networks are hardly the first engineering artifact with downsides – automotive engineers, unintentionally, helped enable suburban sprawl and thousands of traffic deaths; chemical engineers developed DDT; nuclear engineers designed Fukushima; electric power engineering contributed to global climate change. But network engineering had the advantage of being one layer, literally, removed from most of the negative impacts, so engineers building routers could safely, and not incorrectly, blame others for the problems that have become less deniable. But while automotive engineers can directly contribute to the solution, e.g., by building safer and zero-emission cars, the ability of network researchers and engineers to fix the problems that ``their'' artifact has enabled is much harder. No new machine-learning enabled TCP congestion control algorithm for blockchains is going to fix fake news and app-distracted driving.

But the saying attributed to Wernher von Braun about rockets –- “Once the rockets are up, who cares where they come down? That's not my department.”\footnote{Song lyrics by Tom Lehrer} -– may no longer be sufficient in how we train future engineers and scientists, nor how we pitch our technology. We no longer just get to claim credit for the good stuff. A bit of humility would be a good start -– and a good ending for this jeremiad.

\bibliographystyle{IEEEtran}
\bibliography{sample}

\end{document}